\documentclass[lettersize,journal]{IEEEtran}
\usepackage{amsmath,amsfonts}
\usepackage{algorithmic}
\usepackage{algorithm}
\usepackage{array}
\usepackage[caption=false,font=normalsize,labelfont=sf,textfont=sf]{subfig}
\usepackage{textcomp}
\usepackage{stfloats}
\usepackage{url}
\usepackage{verbatim}
\usepackage{graphicx}
\usepackage{cite}
\begin{document}

\title{Stabilizer Formalism for  EAQECCs with Noise  ebits}

\author{~Ruihu Li, Guanmin Guo$^{\ast}$, Yang Liu, and Hao Song
        % <-this % stops a space
\thanks{ Ruihu Li, Guanmin Guo, Yang Liu and Hao Song are with the Fundamentals Department, Air Force
Engineering University, Xi'an, Shaanxi 710051 P. R. China.
Corresponding author: Guanmin Guo. (e-mail: liruihu@aliyun.com;
$^{\ast}$gmguo$\_$xjtukgd@yeah.net;
liu$\_$yang10@163.com; songhao$\_$kgd@163.com)}% <-this % stops a space
\thanks{This work is supported by the National Natural Science Foundation of
China under Grant No.U21A20428.}}

\markboth{IEEE COMMUNICATIONS LETTER,~Vol.~, No.~,}%
{Shell \MakeLowercase{\textit{et al.}}: Stabilizer Formalism for
EAQECCs with Noise  ebits}


\maketitle

\begin{abstract}
We introduce a stabilizer formalism for EAQECCs with noise ebits,
using special subgroups of product groups of two Pauli groups. This
formalism includes the two coding schemes, given by Lai and Brun
(C.Y. Lai and T. A. Brun, PHYSICAL REVIEW A 86, 032319 (2012)), for
EAQECCs with imperfect ebits as special cases. Then two equivalent
formalisms of the formalism are derived in nomenclature of sympletic
geometry and additive codes. We apply this theory to construct some
EAQECCs with noise ebits, and analyze their performance.
\end{abstract}

\begin{IEEEkeywords}
Standard form, stabilizer and normalizer, quantum error-correcting
code, symplectic geometry.
\end{IEEEkeywords}

\section{Introduction}
\IEEEPARstart{Q}{uantum}  error-correcting codes (QECCs) are
necessary to overcome noise in quantum computation and quantum
communication [1,2,3,4]. Quantum stabilizer codes are the most
extensively studied quantum codes, and have the advantage that their
properties can be analyzed using group theory [3,4]. Quantum
stabilizer codes are closely related to classical linear codes, and
can be obtained by the CRSS and CSS code constructions from
self-orthogonal (or dual containing) classical codes [3-7].

Entanglement-assisted quantum error-correcting codes (EAQECCs)
[8-12] are  an extended version of standard QECCs in the literature
[3-4]. EAQECC uses maximally entangled qubits (ebits) that are
shared by the transmitter and receiver. By using these ebits, the
EAQECC is not subject to the dual-containing constraint, and has a
larger minimum distance [9-12]. The entanglement-assisted stabilizer
formalism [9,12] provides a useful framework for constructing
quantum codes.

Most studies on EAQECCs have assumed that errors do not occur on the
shared ebits from the receiver's side because ebits on the
receiver's side do not pass through the transmit channel [8-12].
However, in practice, receiver-side ebits also suffer from errors,
and this reduces the error-correcting capability of the code. The
following works have considered the imperfect (noise) ebits.

Shaw et al.[13] presented an EAQECC that corrects errors on both the
senders qubits and the receivers shared ebits. They showed, for the
first time, that a Steane code is equivalent to a $[[6,1,3;1]]$
EAQECC for correcting a single error on the receiver's (i.e., Bob's)
ebits. Wilde et al. [14] simulated entanglement-assisted quantum
turbo codes when the ebits on Bob's side are imperfect. Their aim
was to analyze the effect of ebit noise on entanglement-assisted
quantum turbo-code performance. Lai and Brun studied a practical
case where errors on the receiver's side can be corrected [15]. They
presented two different schemes to correct errors on the receiver's
side, and showed an equivalent relationship between $[[n,k,d;c]]$
EAQECC and $[[n + c,k,d]]$ stabilizer code. Based on this
equivalence, EAQECCs can correct errors on the ebits of the
receiver's side. In [16], a study of the concatenation of EAQECCs
was initiated, and EAQECCs with noise ebits were discussed.

In this work, we introduce a stabilizer formalism for
 EAQECCs with noise (imperfect)  ebits (EAQECCs-Ne, for short).
This  formalism  provides a new perspective for  EAQECCs-Ne
framework as well as a powerful technique for constructing
EAQECCs-Ne,  and a systematic technique for choosing good EAQECCs in
practice.

The paper is structured as follows. Section II introduces
fundamental notations related to the Pauli group, symplectic space,
and additive codes, which underpin the study of both quantum
error-correcting codes (QECCs) and entanglement-assisted quantum
error-correcting codes (EAQECCs). Building on this, Section III
establishes the stabilizer formalism for EAQECCs involving noisy
entanglement bits (ebits), then present the construction of such
noisy-ebit EAQECCs, followed by a performance comparison with
corresponding optimal quantum codes. Finally, concluding remarks and
future research directions are outlined in the last section.

\section{Preliminaries}

In this section, we introduce some notations of Pauli group,
symplectic space and additive code for QECCs and  EAQECCs, for more
details please see [4,9,12,17-18].

Let ${F}_{2}$ be the  binary field,  ${F}_{2}^{2n}$ be the
$2n$-dimensional symplectic space  whose elements are denoted as $(a
\mid b)$ where $a, b\in {F}_{2}^{n}$. The symplectic inner product
of $(a \mid b)$ and $(a^{'} \mid b^{'})$ is defined to be $((a \mid
b),(a^{'} \mid b^{'}))_{s} =a(b^{'})^{T}-b(a^{'})^{T}$.
 For a subspace $S$ of  ${F}^{2n}_{2}$, its  symplectic
dual is defined as  $S^{\perp s}$ $=\{ (a \mid b)\mid  ((a \mid
b),(a^{'} \mid b^{'}))_{s}=0   \hbox{ for any} (a^{'} \mid b^{'})\in
S\}$. A subspace $S$ is called totally isotropic  if $S\cap S^{\perp
s}=S$ and  non-isotropic  if $S\cap S^{\perp s}=\{0\}$, see [17-18].
A totally isotropic subspace is called an isotropic subspace in
[4,19], and a non-isotropic subspace is called a symplectic subspace
in [4] and an entanglement subspace in [19] respectively. Each
subspace $S$ of ${F}_{2}^{2n}$ can be decomposed as
$S$=$S_{I}$$\oplus S_{E}$, where $S_{I}=$ $S\cap S^{\perp s}$ and $
S_{E}$ is an entanglement subspace [12,17].

Let $F_{4}=\{0, 1, \omega, \varpi\}$  be the  four elements Galois
field  such that $\varpi=1+\omega=\omega^{2}$ and $\omega^{3}=1$.
The conjugation of $x\in F_{4}$ is $\bar x=x^{2}$, and conjugation
transpose of matrix $G$ over $ F_{4}$ is $(\bar G)^{T}=G^{\dagger}$.
The {\it Hermitian inner product} and  {\it trace inner product} of
$ u$, $ v\in$$F_{4}^{n}$ are defined as
 $( u,v)_{h}=u \overline{v}^{T}$$=\sum_{1}^{n}u_{j}v_{j}^{2},$
  $( u,v)_{t}$$=tr( u \overline{v}^{T})=
\sum_{1}^{n}(u_{j}\overline{v_{j}}+\overline{u_{j}}v_{j})$
 $=\sum_{1}^{n}(u_{j}v_{j}^{2}+u_{j}^{2}v_{j}),$
respectively [4,18].

An additive code $\mathcal{C}$ $=(n,2^{m})_{4}$ is an additive
subgroup of $F_{4}^{n}$, the {\it trace dual code} of $\mathcal{C}$
is defined as $\mathcal{C}^{\perp_{t}}=\{u\in F_{4}^{n}\mid
(u,v)_{t}=0 \hbox{ for all } v\in \mathcal{C}\}$, $R_{
t}(\mathcal{C})$ $=\mathcal{C}\cap \mathcal{C}^{\perp_{t}}$ is an
$(n,2^{l})$ additive code called as trace radical of $\mathcal{C}$.
$\mathcal{C}$ is {\it trace self-orthogonal} if
$\mathcal{C}$$\subseteq \mathcal{C}$$^{\perp_{ t}}$, $\mathcal{C}$
is ACD (additive complementary dual) if $R_{ t}(\mathcal{C})$$=\{
0\}$  [20]. Each $\mathcal{C}$ $=(n,2^{m})_{4}$ can be decomposed as
$\mathcal{C}$ $=R_{ t}(\mathcal{C})$$\oplus$$\mathcal{C}$$_{e}$,
$\mathcal{C}$$_{e}$ is an ACD code [17,20]. A linear code
$\mathcal{D}$ $=[n,k]_{4}$ is a subspace of $F_{4}^{n}$ and
$\mathcal{D}$ $=(n,2^{2k})_{4}$.
 The {\it Hermitian
dual code} of $\mathcal{D}$ is defined as $\mathcal{D}^{\perp_{
h}}=\{u\in F_{4}^{n}\mid (u,v)_{h}=0 \hbox{ for all } v\in
\mathcal{D}\}$. $\mathcal{D}$ is {\it  Hermitian self-orthogonal} if
$\mathcal{D}$$\subseteq \mathcal{D}$$^{\perp_{ h}}$, $\mathcal{D}$
is  Hermitian linear complementary dual ( abbreviated to LCD)  if
$R_{ h}(\mathcal{D})$$=\mathcal{D}$$\cap \mathcal{D}$$^{\perp_{ h}}$
$=\{ 0\}$. Each $\mathcal{D}$ $=[n,k]_{4}$ can be decomposed as
$\mathcal{D}$ $=R_{ h}(\mathcal{D})$$\oplus$$\mathcal{D}$$_{e}$,
$\mathcal{D}$$_{e}$ is an Hermitian LCD code [12,20].

There is an isometry map $\phi$ from $F_{2}^{2n}$ to $F_{4}^{n}$ as
 $\phi((a\mid b))=\omega a+\varpi b \in F_{4}^{n}$
for $v=(a\mid b)\in F_{2}^{2n}$, see [4,18]. If $S$ is a
$m$-dimensional subspace of $F_{2}^{2n}$, $\mathcal{C}=\phi(S)$ is
an $(n, 2^{m})$ additive code, $S$=$S_{I}$$\oplus S_{E}$ implies
$\mathcal{C}$$=R_{ t}(\mathcal{C})$$\oplus$$\mathcal{C}$$_{e}$,
where $R_{ t}(\mathcal{C})=\phi(S_{I})$ and $\mathcal{C}$$_{e}$
$=\phi( S_{E})$ [12,18].

Let $\mathcal{P}$ $=\{ I=I_{2}, X, Y, Z  \}$ be the set of Pauli
operators, where
$$ I= \left(
\begin{array}{cccccc}
1&0\\
0&1\\
\end{array}
\right), X= \left(
\begin{array}{cccccc}
0&1\\
1&0\\
\end{array}
\right),$$

$$ Z= \left(
\begin{array}{cccccc}
1&0\\
0&1\\
\end{array}
\right), Y= \left(
\begin{array}{cccccc}
0&-i\\
i&0\\
\end{array}
\right).$$

The $n$-fold Pauli group is $\mathcal{P}$$_{n} $
$=\{i^{a}M_{1}\otimes M_{2}\otimes\cdots M_{n}: a=0,1,2,3, M_{j}\in$
$\mathcal{P} \}$. Every element $e\in $ $\mathcal{P}$$_{n} $ can be
written uniquely in the form $e=i^{l}X(a)Z(b)$, where $l=0,1,2,3$,
$a,b\in $ ${F}_{2}^{n}$ [4,18]. Then $\mathcal{G}$$_{n}
$$=\mathcal{P}$$_{n} $ $/\{ \pm I_{2^{n}}, \pm
 iI_{2^{n}}   \}$ is isometry
isomorphism to  the
 symplectic space ${F}_{2}^{2n}$ under the
map  $\tau(i^{l}X(a)Z(b))=(a \mid b)$ [4,12]. If $\mathcal{A}$ is a
subgroup of $\mathcal{G}$$_{n}$, then $\tau(\mathcal{A})$ is a
subspace of ${F}_{2}^{2n}$. $\mathcal{A}$ is an Abelian subgroup  if
and only if its image $\tau(\mathcal{A})$ is a
 totally isotropic subspace.
 If $\tau(\mathcal{A})$ is a
 totally isotropic subspace of ${F}_{2}^{2n}$, $\mathcal{A}$ is called an isotropic
subgroup of $\mathcal{G}$$_{n}$. If $\tau(\mathcal{A})$ is a
non-isotropic subspace of ${F}_{2}^{2n}$, $\mathcal{A}$ is called a
symplectic subgroup of  $\mathcal{G}$$_{n}$ in [12] and an
entanglement subgroup in [13], respectively. For a subgroup
$\mathcal{S}$ of  $\mathcal{P}$$_{n}$, let
$\mathcal{N(\mathcal{S})}$ be its normalizer. According to [21],
$\mathcal{S}$ has decomposition  $\mathcal{S}$
$=\mathcal{S}$$_{I}$$\times \mathcal{S}$$_{E}$, where
$\mathcal{S}$$_{I}=$ $\mathcal{S}$$\cap\mathcal{N(\mathcal{S})}$ is
an isotropic subgroup, $\mathcal{S}$$_{E}$ is an entanglement
subgroup.

According to [3,4] and [9, 12], we have the following theorem.

{\bf Theorem 2.1$^{[3,4]}$} Let $\mathcal{S}$ be an Abelian subgroup
of $\mathcal{P}$$_{n}$ of size $2^{m}$  and
$\mathcal{N(\mathcal{S})}$ be its normalizer. If $-I_{2^{n}}$
$\not\in$ $\mathcal{S}$, then $\mathcal{S}$ fix a QECC
$\mathcal{Q}$$=[[n,k,d]]$, where $k=n-m$, $d=min \{wt(g) \mid g\in
\mathcal{N(\mathcal{S})}\setminus$ $\mathcal{S}\}$, $\mathcal{S}$ is
called the stabilizer of $\mathcal{Q}$.

{\bf Theorem 2.2$^{[9, 12]}$} Let $\mathcal{S}$ be a subgroup of
$\mathcal{P}$$_{n}$ of size $2^{m}$, $\mathcal{S}$$_{I}$ be an
isotropic subgroup of size $2^{l}$ and $\mathcal{S}$$_{E}$ be an
entanglement subgroup of size $2^{2c}$. If $\mathcal{S}$
$=\mathcal{S}$$_{I}$$\times \mathcal{S}$$_{E}$, then $\mathcal{S}$
can be extended into an Abelian subgroup $\mathcal{\tilde{S}}$ of
$\mathcal{P}$$_{n+c}$ with $c$ maximally entangled  pairs.
$\mathcal{\tilde{S}}$ fix an EAQECC
$\mathcal{Q}^{ea}$$=[[n,k,d_{ea};c]]$, where $k=n+c-m=n-c-l$,
$d_{ea}=min \{wt(g) \mid g\in \mathcal{N(\mathcal{S})}\setminus$
$\mathcal{S}$$_{I}\}$. $\mathcal{S}$ is called the EA- stabilizer of
$\mathcal{Q}$$^{ea}$.

Suppose  $\mathcal{Q}$ is an $[[n+c,k,d]]$ QECC with stabilizer
$\mathcal{S}$, $\mathcal{Q}$$^{ea}$ is an $[[n,k,d;c]]$ EAQECC with
EA-stabilizer $\mathcal{S'}$. If  $\mathcal{S'}$ is obtained from
$\mathcal{S}$ by puncturing some $c$ coordinates, we say
$\mathcal{Q}$$^{ea}$ is equivalent to $\mathcal{Q}$.

\section{Main Results}

\subsection{Stabilizer formalism for EAQECCs-Ne}

We will use group theory to describe EAQECCs with noise ebits
(EAQECCs-Ne), establish stabilizer formalism for EAQECCs-Ne.

Suppose that Alice (the sender) uses a
$\mathcal{Q}$$^{ea}$$=[[n,k,d_{ea};c]]$ to send $2^{k}$ messages
through noise  channel $\mathcal{N}$$_{A}$, such an EAQECC has  EA-
stabilizer $\mathcal{S}$. Suppose also the receiver Bob use a
quantum code $\mathcal{Q}$$^{b}$$=[[m,k^{b},d_{b}]]$ with stabilizer
$\mathcal{S}$$^{b}$ to protect his $c$ ebits, hence there holds
$k^{b}\geq c$, assume this storage system of $m$ qubits is a less
noise channel $\mathcal{N}$$_{B}$. The error group act on Alice's
side is Pauli group $\mathcal{P}$$_{n}$,
 the error group act on  Bob's side is Pauli group $\mathcal{P}$$_{m}$, we denote this
 $\mathcal{P}$$_{m}$ as
$\mathcal{P}$$^{b}_{m}$. Let $\mathcal{G}=$
$\mathcal{P}$$_{n}$$\ast$$\mathcal{P}$$^{b}_{m}$= $\{ (g,g')$:
$g\in\mathcal{P}$$_{n}$, $g'\in\mathcal{P}$$^{b}_{m}$$\}$ be product
group of $\mathcal{P}$$_{n}$ and $\mathcal{P}$$^{b}_{m}$. We will
use subgroups of  $\mathcal{G}$ to describe combination EAQECC
$\mathcal{Q}$$^{ea}_{c}$$=$$\mathcal{Q}$$^{ea}$+$\mathcal{Q}$$^{b}$
$=[[n,k,d_{ea} ;c]]$$+[[m,k^{b},d_{b}]]$.

Now, we discuss  Abelian subgroup $\mathcal{S}$$^{b}$ of
$\mathcal{P}$$^{b}_{m}$ such that  $\mathcal{S}$$^{b}$ stabilizes an
$\mathcal{Q}$$=[[m,k^{b}]]$ QECC with $k^{b}\geq c$.

{\bf Definition 3.1} Let $\mathcal{S}$ be a subgroup of
$\mathcal{P}$$_{n}$, $\mathcal{S}$$^{b}$ be an Abelian subgroup of
$\mathcal{P}$$^{b}_{m}$ and $-I_{2^{m}}$
$\not\in$$\mathcal{S}$$^{b}$.

(1) $\mathcal{S}$$^{b}$ match $\mathcal{S}$ if $\mathcal{S}$
EA-stabilizes a $\mathcal{Q}$$^{ea}$$=[[n,k;c]]$ and
$\mathcal{S}$$^{b}$ stabilizes an $[[m,k^{b}]]$ code with $k^{b}\geq
c$.  ($\mathcal{S}$, $\mathcal{S}$$^{b}$) is a matching subgroup of
$\mathcal{P}$$_{n}$ $\ast \mathcal{P}$$^{b}_{m}$ if
$\mathcal{S}$$^{b}$ match $\mathcal{S}$.

(2) If $\mathcal{S}$$^{b}$ stabilizes an $[[m,k^{b},d_{b}]]$ code
with $d_{b}\geq 3$, a matching subgroup ($\mathcal{S}$,
$\mathcal{S}$$^{b}$) is called a faithful matching subgroup.

(3) $\mathcal{S}$$^{b}$ properly match $\mathcal{S}$ if
$\mathcal{S}$ EA-stabilizes a $\mathcal{Q}$$^{ea}$$=[[n,k;c]]$ and
$\mathcal{S}$$^{b}$ stabilizes an $[[m,c ]]$ code. ($\mathcal{S}$,
$\mathcal{S}$$^{b}$) is a properly matching subgroup of
$\mathcal{P}$$_{n}$$\ast \mathcal{P}$$^{b}_{m}$ if
$\mathcal{S}$$^{b}$  properly match $\mathcal{S}$.

{ \bf Proposition 3.1} (1) Let $M_{s}=(4^{s}-1)/3$ for $s\geq 2$,
$M_{s-1}< c\leq M_{s}$. Then there is an $[[m,c,3]]$ quantum code
for some $m\leq 2M_{s}$.

 (2)  If $\mathcal{S}$
EA-stabilizes a $\mathcal{Q}$$^{ea}$$=[[n,k;c]]$, then there is an
 $\mathcal{S}$$^{b}$ in $\mathcal{P}$$^{b}_{m}$ faithful matching
 $\mathcal{S}$ for some $m\leq 2M_{s}$, where $s=\lfloor
 log_{2}3c/2\rfloor$.

{\bf Proof} According to [22,23], if $M_{s-1}< c\leq M_{s}-5$ or $
c= M_{s+1}$, there is  an $[[m=c+2s,c,3]]$ quantum code; if
$M_{s}-5\leq c\leq  M_{s}$, there is an $[[m=c+2(s+1),c,3]]$ quantum
code. Thus,  (1) holds.

 (2) follows from  (1).

{\bf Theorem 3.2} Let $\mathcal{S}$ $=\mathcal{S}$$_{I}$$\times
\mathcal{S}$$_{E}$ be  a subgroup of $\mathcal{P}$$_{n}$,
$\mathcal{S}$$_{I}$ be an isotropic subgroup of size $2^{l}$ and
$\mathcal{S}$$_{E}$ be an entanglement subgroup of size $2^{2c}$,
$-I_{2^{m}}$ $\not\in$ $\mathcal{S}$$^{b}$ and $\mathcal{S}$$^{b}$
be an Abelian subgroup of $\mathcal{P}$$^{b}_{m}$ of size
$2^{m-k^{b}}$. If $k^{b}\geq c$, $M_{s-1}\leq m\leq 2M_{s}$, then
there is a combination EAQECC
$\mathcal{Q}$$^{ea}_{c}$$=$$[[n,k,d_{ea} ;c]]$ $+[[m,k^{b},d^{b}\geq
3]]$, where $s=\lfloor log_{2}3k^{b}/2\rfloor$, $k=n-c-l$.

{\bf  Proof} According to Theorem 3.2,  $\mathcal{S}$ EA-stabilizes
an $[[n,k,d_{ea};c]]$. $\mathcal{S}$$^{b}$ stabilizes an
$[[m,k^{b},d^{b} ]]$ QECC, this code can protect the receiver's $c$
ebits when  $k^{b}\geq c$. Hence, one can obtain a combination
EAQECC as desired.

{\bf  Notation 3.1} (1) If  $\mathcal{S}$   EA-stabilizer a
$\mathcal{Q}$$^{ea}$$=[[n,k,d_{ea};c]]$ and $\mathcal{Q}$$^{ea}$
equivalent to an $[[n+c,k,d]]$ QECC, then choose $\mathcal{S}$$^{b}$
$=\{ I_{2^{m}} \}$, the combination code
$\mathcal{Q}$$^{ea}_{c}$$=$$[[n,k,d_{ea} ;c]]$ $+[[c,c,1]]$ has the
same function as $\mathcal{Q}$$^{ea}$$=$$[[n,k,d_{ea} ;c]]$ EAQECC.

(2) If  $\mathcal{S}$   EA-stabilizes a
$\mathcal{Q}$$^{ea}$$=[[n,k,d_{ea};c]]$ and $\mathcal{Q}$$^{ea}$ is
not equivalent to an $[[n+c,k,d]]$ QECC, choose a
$\mathcal{S}$$^{b}$ with size $2^{m-c}$ and match  $\mathcal{S}$,
then $\mathcal{Q}$$^{ea}_{c}$$=$$[[n,k,d_{ea} ;c]]$ $+[[m,c]]$ as
given in [15].

Thus, the two schemes  of [15] for EAQECC with imperfect ebits  are
special cases of our formalism.

It is very hard to construct EAQECCs-Ne using the framework of
Theorem 3.2. Using  relationships  among $\mathcal{P}$$_{n}$,
${F}_{2}^{2n}$ and ${F}_{4}^{n}$, two equivalent formalisms of
Theorem 3.2 can be given as follows.

{\bf  Theorem 3.3} Suppose a subspace $S$   of ${F}_{2}^{2n}$ is
decomposed as $S$=$S_{I}$$\oplus S_{E}$, where $S_{I}=$ $S\cap
S^{\perp s}$ and $ S_{E}$ is an entanglement subspace, $ S^{b}$ is
an  isotropic subspace of ${F}_{2}^{2m}$, $|S_{I}|=2^{l}$ and
$|S_{E}|$$=2^{2c}$ and $|S^{b}|= 2^{r}$. If  $c\leq m-r$,
$M_{s-1}\leq m\leq 2M_{s}$, then there is a combination EAQECC
$\mathcal{Q}$$^{ea}_{c}$$=$$[[n,k,d_{ea} ;c]]$ $+[[m,k^{b}]]$, where
$k=n-c-l$, $k^{b}=m-r\geq c$ and $s=\lfloor log_{2}3k^{b}/2\rfloor$.

{\bf  Theorem 3.4} Suppose an additive code $\mathcal{C}$
 is decomposed as $\mathcal{C}$ $=R_{
t}(\mathcal{C})$$\oplus$$\mathcal{C}$$_{e}$, $\mathcal{C}$$_{e}$ is
an ACD code, $\mid R_{ t}(\mathcal{C})\mid=2^{l}$ and $\mid
\mathcal{C}$$_{e}\mid$$=2^{2c}$. If $\mathcal{C}$$^{b}$
$=(m,2^{r})_{4}$ is a trace self-orthogonal code with $c\leq
m-r=k^{b}$, $M_{s-1}\leq m\leq 2M_{s}$, then there is a combination
EAQECC $\mathcal{Q}$$^{ea}_{c}$$=$$[[n,k,d_{ea} ;c]]$
$+[[m,k^{b}]]$, where $k=n-c-l$,  and $s=\lfloor
log_{2}3k^{b}/2\rfloor$.

When the two  additive codes in Theorem 3.4 are all linear codes, we
have the following results.

{\bf Corollary 3.2} Suppose a linear code $\mathcal{D}$$=[n,u]_{4}$
 is decomposed as $\mathcal{D}$ $=R_{
h}(\mathcal{D})$$\oplus$$\mathcal{D}$$_{e}$, $\mathcal{D}$$_{e}$ is
a  Hermitian LCD code, $r=dim R(\mathcal{C})$. If
$\mathcal{D}$$^{b}$ $=[m,v]_{4}$ is a  Hermitian self-orthogonal
code with $c=u-r\leq k^{b}= m-2v$,  $M_{s-1}\leq m\leq 2M_{s}$, then
there is a combination EAQECC
$\mathcal{Q}$$^{ea}_{c}$$=$$[[n,k,d_{ea} ;c]]$$+[[m,k^{b}]]$, where
$k=n-c-2r$, $c= u-r$,  $s=\lfloor log_{2}3k^{b}/2\rfloor$.

\subsection{Construction of EAQECCs-Ne}

Using Theorem 3.4, especially Corollary 3.2, one can obtain
 EAQECCs-Ne. If $[[n,k,d_{ea} ;c]]$ is equivalent to an $[[n+c,k,d]]$
 QECC, such an EAQECC-Ne has the same ability of correcting error as
$[[n+c,k,d]]$ QECC.

{ \bf Proposition 4.1} Let $\mathcal{C}$$=[N,u]_{4}$ be Hermitian
self-orthogonal code with dual distance $d$, $\mathcal{C}$
stabilizes a pure $\mathcal{Q}$$=[[N,N-2u,d]]$. Hence there are
$[[n,k=N-2u,d;c]]$ for $n=N-c$ and $0<c\leq u$. All these EAQECCs
are equivalent to QECCs.

{ \bf Example 4.1} For $u=6$, let $N=13,14,15,18$, there are $[N,6]$
SO codes with dual distance $d^{\perp_{h}}=5$. Hence there are
$[[n,k=N-12,5;c]]$ for $n=N-c$ and $0<c\leq u$. All these EAQECCs
can correct $t\leq 2$ errors on message bits and ebits.

In the following, we fix on constructing
$\mathcal{Q}$$^{ea}_{c}$$=$$[[n,k,d_{ea} ;c]]$$+[[m,c]]$, with
$[[n,k,d_{ea} ;c]]$ does not equivalent to QECC. Firstly, we present
some EAQECCs-Ne from known EAQECCs in [12,23].

{\bf Proposition 4.2} (1) If $m\geq 2$, there are
 EAQECCs-Ne with parameters $[[4m,1,2m+1 ;1]]$$+[[5,1,3]]$,
 $[[4m+1,1,2m+3 ;4]]$$+[[10,4,3]]$,
 $[[4m+2,1,2m+3 ;3]]$$+[[8,3,3]]$,
$[[4m+3,1,2m+3 ;2]]$$+[[8,2,3]]$.

(2) There are EAQECCs-Ne with parameters $[[7,2,5
;5]]$$+[[11,5,3]]$, $[[8,2,5 ;4]]$$+[[10,4,3]], [[9,2,5;3]]$
$+[[8,3,3]]$, $[[10,2,6 ;4]]$$+[[10,4,3]]$, $[[9,3,6
;6]]$$+[[12,6,3]]$, $[[13,3,9 ;10]]$$+[[16,10,3]]$, $[[12,4,7
;8]]$$+[[14,8,3]]$.

{\bf  Proof} (1) According to [24], there are EAQECCs with
parameters $[[4m,1,2m+1 ;1]]$, $[[4m+1,1,2m+3 ;4]]$, $[[4m+2,1,2m+3
;3]]$ and $[[4m+3,1,2m+3 ;2]]$. Using known $[[m,c,3]]$ codes given
in [4,23], one can derive (1) holds.

(2) Similar to discussion of (1), from EAQECCs obtained in [12], we
know (2) holds.

{\bf Theorem 4.3} Let $\mathcal{C}$ $=(n,2^{l})$, $\mathcal{C'}$
$=(n,2^{2k})$, $\mathcal{E}$$=(n,2^{2k},d_{2})$,
 $\mathcal{D}=$ $\mathcal{C}+\mathcal{C'}$$=(n,2^{l+2k},d_{1})$.
Suppose $G'$ and $E$ are generator matrices of $\mathcal{C'}$ and
$\mathcal{E}$, respectively. If  $\mathcal{C}$ $\subseteq
\mathcal{D}$ $ \subseteq\mathcal{C}^{\perp t}$, $(G'\mid E)$
generates an ACD code, then there is an $[[n+m,k, d ; c]]$
EAQECCs-Ne, where $d\geq d_{1}+d_{2}$, $c=n+m-l-k$.

{\bf Proof} Let $G$ be a generator matrix of $\mathcal{C}$,
construct
$$G_{l+2k,n+m}=\left(
\begin{array}{lc}
G \mid {\bf 0_{l,m}} \\
G'\mid  E \\
\end{array}
 \right).$$

Then $G_{l+2k,n+m}$ generates an $(n+m,2^{l+2k},d_{1})$ code
$\mathcal{M}$ with decomposition   $\mathcal{M}$$=R_{
t}(\mathcal{M})$$\oplus$$\mathcal{M}$$_{e}$, $(G'\mid E)$ generates
$\mathcal{M}$$_{e}=(n+m,2^{2k},d)$ and $d\geq d_{1}+d_{2}$. Using
$\mathcal{M}$$^{\perp_{ t}}$ as EA-stabilizer, one can get an
$[[n+m,k, d ; c]]$ EAQECC  with $c=n+m-l-k$.

{\bf Corollary 4.4} There are EAQECCs-Ne with parameters\\
$[[14,2,7;4]]$$+[[10,4,3]]$,  $[[15,2,8;5]]$$+[[12,6,3]], [[16,2,7
;4]]$ $+[[10,4,3]]$, $[[17,2,8 ;5]]$$+[[12,6,3]], [[20,2,9
;4]]$$+[[10,4,3]]$, $[[21,2,9 ;5]]$$+[[12,6,3]]$, $[[30,3,13
;9]]$$+[[15,9,3]], [[31,3,13 $\\
$;8]]+[[15,9,3]]$, $[[33,3,13;6]]$$+[[12,6,3]]$.

{\bf  Proof} Case 1. Let $\mathcal{D}$$_{1}$ $=(12,2^{8+2\times
2},6)$, $=(12,2^{12},6)$ be the additive self-dual code given in
[4], $ \mathcal{E}$$_{1}$$=(2,2^{4},1)$ $ \mathcal{E}$$_{2}$
$=(3,2^{4},2)$ be ACD codes. Then one can derives $[[14,2,7;4]]$ and
$[[15,2,8;5]]$ from  Theorem 4.3.

Case 2. Let $ \mathcal{D}$$_{2}$ $=(14,2^{10+2\times 2},6)$
self-dual code given in [4]. Similar to Case 1, $[[16,2,7;4]]$ and
$[[17,2,8;5]]$ can be deduced from  Theorem 4.3.

Case 3. Let $ \mathcal{D}$$_{3}$ $=(18,2^{14+2\times 2},8)$
self-dual code given in [4]. Similar to Case 1, $[[20,2,9;4]]$ and
$[[21,2,10;5]]$ can be deduced from  Theorem 4.3.

Case 4. Let $ \mathcal{D}$$_{4}$ $=(30,2^{24+2\times 3},12)$
self-dual code given in [4].  From $ \mathcal{D}$$_{4}$, one can
obtain   $ \mathcal{D}$$_{5}$$=(27,2^{18+2\times 3},12)$,  $
\mathcal{D}$$_{6}$, $=(28,2^{20+2\times 3},12)$ self-orthogonal
codes. Choose $\mathcal{E}_{3}=(3,2^{6},1)$ ACD code, similar to
Case 1, we can construct $[[30,3,13 ;9]]$$+[[15,9,3]]$, $[[31,3,13
;8]]$$+[[15,9,3]]$, $[[33,3,13 ;6]]$$+[[12,6,3]]$ from $
\mathcal{D}$$_{5}$, $\mathcal{D}$$_{6}$, $ \mathcal{D}$$_{4}$ and $
\mathcal{E}$$_{3}$.

To sum up, this conclusion is valid.

\subsection{Performance of EAQECCs-Ne}

In this section, we compare performance of some of our $[[n,k, d ;
c]]+[[m,c,d^{b}]]$ EAQECC-Ne with optimal $[[N,k, d ]]$ code. In
[15],  Lai and Brun compare performance of some small codes by
computing their channel fidelity. They also pointed out that it is
difficult to get true channel fidelity $F( \mathcal{C})$ of a code $
\mathcal{C}$, $F( \mathcal{C})$ has good approximation
$P(\mathcal{C})$ = $Pr(\{\hbox{errors of weight less than or equal
to}  \frac{\lfloor d-1\rfloor}{2} \})$, where $d$ is the minimum
distance of the quantum code $ \mathcal{C}$ [15]. We will use
$P(\mathcal{C})$ to compare performance of our $[[n,k, d ;
c]]+[[m,c,d^{b}]]$ EAQECC-Ne with those of optimal $[[N,k, d ]]$
code.

Suppose that Alice send quantum information to Bob through noise
channel $\mathcal{N}$$_{A}$, Bob use a quantum standard stabilizer
code (less noise  channel $\mathcal{N}$$_{B}$) to protect his $c$
ebits. Let $p_{a}$ and $p_{b}$ be the depolarizing rate of
$\mathcal{N}$$_{A}$ and $\mathcal{N}$$_{B}$, respectively.

For a $\mathcal{C}=[[N,k,d]]$ used on noise  channel
$\mathcal{N}$$_{A}$. Let $t=\frac{\lfloor d-1\rfloor}{2}$, its
$P(\mathcal{C})$ is
\begin{align*}
P(\mathcal{C})&=Pr(\{\hbox{errors of weight less than or equal to}
\frac{\lfloor d-1\rfloor}{2} \})\\
&=\sum_{i=0}^{t}(1-p_{a})^{N-i} p_{a}^{i}\binom {N}{i}.
\end{align*}

For an EAQECC-Ne $\mathcal{D}=[[n,k, d ; c]]+[[m,c,d^{b}]]$
$=\mathcal{D}^{ea}+\mathcal{D}^{b}$, let $t=\frac{\lfloor
d-1\rfloor}{2}$ and $t^{b}=\frac{\lfloor d^{b}-1\rfloor}{2}$. The
code $\mathcal{D}^{b}=[[m,c,d^{b}]]$ is used to protect $c$ ebits,
its $P(\mathcal{D}^{b})$$= \sum_{i=0}^{t^{b}}(1-p_{b})^{m-i}
p_{b}^{i}\binom {m}{i}$. Once Bob decode $\mathcal{D}^{b}$
correctly, he can then decode $\mathcal{D}^{ea}=[[n,k, d ; c]]$
correctly with   probability
$P(\mathcal{D}^{ea})$$=\sum_{i=0}^{t}(1-p_{a})^{n-i} p_{a}^{i}\binom
{n}{i}$. Hence $\mathcal{D}=[[n,k, d ; c]]+[[m,c,d^{b}]]$ has
$P(\mathcal{D})$$=P(\mathcal{D}^{ea})P(\mathcal{D}^{b})$.

If $P(\mathcal{D})$$>P(\mathcal{C})$, then $\mathcal{D}=[[n,k, d ;
c]]+[[m,c,d^{b}]]$ has better  performance than $\mathcal{C}=[[N,k,
d ]]$.

{\bf  Example 5.1} The smallest optimal $[[N,1,7]]$ has $N=17$, now
we compare performance of the $[[12,1,7 ;1]]+[[5,1,3]]$ quantum code
with those of $[[17,1,7]]$. Our $[[12,1,7 ;1]]$$+[[5,1,3]]$ has
better performance than $[[17,1,7]]$ when $ p_{b}$ $=p_{a}\lambda$
for suitable small $\lambda$ as shown in TABLE I.

%From the perspective of channel fidelity,

\begin{table}[!t]
\caption{Performance comparison between $[[12,1,7 ;1]]$$+[[5,1,3]]$
and $[[17,1,7]]$. \label{tab:table1}} \centering
\begin{tabular}{|c|c|c|c|c|c|c|c|c|c|c|c|}
\hline
$p_{a}$ &  max $ \{p_{b}\}$  &  $\lambda$ & $P(\mathcal{D})$ &  $P(\mathcal{C})$ \\
\hline
0.0100  &  0.0013 &   0.1288  &    0.999979&      0.999979 \\
\hline
0.0199&    0.0049  &  0.2440&      0.999698   &   0.999697\\
\hline
0.0298  &  0.0103  &  0.3472   &   0.998630  &    0.998626\\
\hline
0.0397  &  0.0175   & 0.4403 &     0.996102  &    0.996101\\
\hline
0.0496 &   0.0260 &   0.5235   &   0.991446 &     0.991444\\
\hline
0.0595  &  0.0356 &   0.5976 &     0.984073&      0.984047 \\
\hline
0.0694 &   0.0461  &  0.6647   &   0.973462   &   0.973424\\
\hline
0.0793   & 0.0575 &   0.7248   &   0.959300 &     0.959232\\
\hline
0.0892  &  0.0695 &   0.7789  &    0.941380 &     0.941284\\
\hline
0.0991  &  0.0820 &   0.8280  &    0.919611 &     0.919544\\
\hline
0.1090  &  0.0950 &   0.8721   &   0.894132  &    0.894109\\
\hline
0.1189  &  0.1083  &  0.9111  &    0.865248 &     0.865197 \\
\hline
0.1288 &   0.1219 &   0.9462   &   0.833204  &    0.833122 \\
\hline
0.1387 &   0.1355&    0.9773  &    0.798477  &    0.798272 \\
\hline
0.1486  &  0.1485  &  0.9993  &    0.762866  &    0.761089  \\
\hline
0.1585  &  0.1584  &  0.9994 &     0.730771  &    0.722046 \\
 \hline
 \hline
\end{tabular}
\end{table}

{\bf Example 5.2} The smallest optimal $[[N,4,7]]$ has $N=25$, \\
now we compare performance of the $[[12,4,7 ;8]]$$+[[14,8,3]]$
quantum code with those of $[[25,4,7]]$. Our $[[12,4,7
;8]]$$+[[14,8,3]]$ code are better than $[[25,4,7]]$ when $ p_{b}$
$=p_{a}\mu$ for suitable small $\mu$ as shown in TABLE II.

{\bf Example 5.3} The smallest optimal $[[N,3,9]]$ has $N=27$, \\
now we compare performance of the  $[[13,3,9 ;10]]$ $+[[16,10,3]]$
quantum code with those of $[[27,3,9]]$. Our $[[13,3,9
;10]]$$+[[16,10,3]]$ code are better than $[[27,3,9]]$ when $ p_{b}$
$=p_{a}\nu$ for suitable small $\nu$ as shown in TABLE III.

\begin{table}[!t]
\caption{Performance comparison between $[[12,4,7 ;8]]$$+[[14,8,3]]$
and $[[25,4,7]]$. \label{tab:table1}} \centering
\begin{tabular}{|c|c|c|c|c|c|c|c|c|c|c|c|}
\hline
$p_{a}$ &  max $ \{p_{b}\}$  &  $\mu$ & $P(\mathcal{D})$ &  $P(\mathcal{C})$ \\
\hline
0.0100 &  0.0011&   0.1060  &   0.999894  &   0.999893\\
\hline
0.0199   & 0.0039  & 0.1962  &   0.998587 &    0.998581\\
\hline
0.0298 &  0.0082 &  0.2744  &  0.993983  &   0.993959\\
\hline

0.0397 &  0.0136 &  0.3425 &    0.983974 & 0.983891\\
\hline

0.0496  & 0.0200  & 0.4036  &   0.966818  &   0.966773\\
\hline

0.0595 &  0.0272 &   0.4567  &   0.941915 &    0.941738\\
\hline

0.0694  & 0.0350   & 0.5048 &   0.908826  &   0.908666\\
\hline

0.0793 &  0.0434 &  0.5479 &    0.868170  & 0.868071\\
\hline

0.0892 & 0.0523 &  0.5859 &    0.821227  &   0.820950\\
\hline

0.0991  &0.0615  & 0.6210&      0.768705  &   0.768609\\
\hline

0.1090  & 0.0711  & 0.6521 &   0.712740  &   0.712515\\
\hline \hline
\end{tabular}
\end{table}

\begin{table}[!t]
\caption{Performance comparison between $[[12,1,7 ;1]]$$+[[5,1,3]]$
and $[[17,1,7]]$. \label{tab:table1}} \centering
\begin{tabular}{|c|c|c|c|c|c|c|c|c|c|c|c|}
\hline
 $p_{a}$ &  max $ \{p_{b}\}$  &  $\nu$ & $P(\mathcal{D})$ &  $P(\mathcal{C})$ \\
  \hline
0.0100 &  0.0002  & 0.0228 &    0.999994 &    0.999993 \\
\hline
0.0199 &  0.0012 &  0.0598&     0.999829 &    0.999825 \\
\hline
0.0298 &  0.0030 &  0.1008  &   0.998923  &   0.998904 \\
\hline
0.0397 &  0.0057  & 0.1438   &  0.996197   &  0.996172 \\
\hline
0.0496  & 0.0092 &  0.1858  &   0.990374  &   0.990303 \\
\hline
0.0595 &  0.0135  & 0.2268  &   0.980094 &    0.979958 \\
\hline
0.0694  & 0.0185  & 0.2669 &    0.964105 &    0.964009 \\
\hline
0.0793  & 0.0242 &  0.3049 &    0.941739 &    0.941687 \\
\hline
0.0892 &  0.0304 &  0.3410   &  0.912669  &   0.912658 \\
\hline
0.0991 &  0.0371  & 0.3740  &   0.877536  &   0.877031 \\
\hline
0.1090 &  0.0444 &  0.4070   &  0.835335   &  0.835306 \\
\hline
0.1189  & 0.0520 &  0.4371  &   0.788506  &   0.788308 \\
\hline
0.1288 &  0.0599  & 0.4651   &  0.737672   &  0.737087 \\
\hline
0.1387 &  0.0683 &  0.4921  &   0.683293   &  0.682827 \\
\hline
0.1486  & 0.0768 &  0.5172 &    0.627425  &   0.626755 \\
\hline \hline
\end{tabular}
\end{table}

\section{Conclusion}
In this paper, we have developed a stabilizer description of EAQECCs
with noise ebits. This stabilizer formalism for EAQECCs-Ne
generalizes the known schemes given in [15] for EAQECCs with noise
ebits, transforms the problem of finding EAQECCs-Ne into the problem
of finding two additive codes with some connections. We constructed
some EAQECCs-Ne with very good parameters. Finally, we have
illustrated that some of our EAQECCs-Ne have better performance than
optimal stabilizer codes with the same error correcting ability. An
important issue which remains open is to optimize  parameters of
EAQECCs-Ne $[[n,k, d ; c]]+[[m,c,d^{b}]]$ for given $k, d$ and $c$.
We plan to pursue this investigation in the future and invite others
to do so as well.

\end{document}